# Stream-K++: Adaptive GPU GEMM Kernel Selection and Scheduling for AI using Bloom Filters

Harisankar Sadasivan[1*][0000-0002-2832-458X], Muhammed Emin Ozturk[2][0009-0001-3400-4195], Muhammad Osama[2][0000-0003-1616-6817], Chris Millette[2][0009-0003-2385-245X], Astha Rai[2][0009-0007-7611-293X], Maksim Podkorytov[2][0000-0003-3312-8038], John Afaganis[2][0009-0006-9577-099X], Carlus Huang[2][0009-0009-3048-7642], Jing Zhang[3*][0000-0001-8114-1080], and Jun Liu[3*][0009-0003-4945-1240]

[1] University of Washington, Seattle, USA
hsadasiv@uw.edu
[2] Advanced Micro Devices, Inc. USA / Canada / China
{muhammed.ozturk, muhammad.osama, chris.millette, astha.rai, maksim.podkorytov, john.afaganis, carlus.huang}@amd.com
[3] Meta Platforms, Inc., Austin / Menlo Park, USA
{jizhan, junliume}@meta.com

Abstract. General matrix multiplication (GEMM) operations are the fundamental building blocks of computational domains including artificial intelligence (AI). As GPU architectures evolve and high-performance AI becomes increasingly important, optimizing GEMM performance becomes a fundamental problem that needs to be addressed. This paper introduces Stream-K++, an enhancement to the promising Stream-K GEMM scheduling algorithm for workload balancing. We expand StreamK's scheduling policies from three to seven and implement an efficient solution selection mechanism using Bloom filters. Our approach rapidly eliminates up to ∼95.8% of unsuitable configurations while maintaining a 100% true-negative rate.

Implemented using the AMD Composable Kernel library and evaluated on AMD Instinct™ MI250X GPUs, Stream-K++ demonstrates significant performance gains (up to ∼43%) in select scenarios. It remains competitive (within 20% of optimal) for ∼60-97.6% of problem sizes. Our flexible framework, implemented in the Open-sieve C++ library, allows for easy adaptation to new problem sizes, scheduling policies, or additional tuning parameters, paving the way for future optimizations in GPU-based GEMM operations.

---

* H.S, J.Z and J.L contributed to this work while they were employed full-time at AMD.

## 1 Introduction

General matrix multiplication (GEMM) operations are fundamental to a wide range of computational tasks, from machine learning [31] and scientific simulations [13] to computer graphics [25] and signal processing [30]. The importance of GEMMs in artificial intelligence (AI) has grown exponentially with the rise of transformer models [32,8] in natural language processing. For example, the BERT model [12], which revolutionized language understanding, relies heavily on matrix multiplications in its self-attention mechanism. Similarly, GPT-3 [19], with its 175 billion parameters, performs an enormous number of GEMM operations during both training and inference. Beyond transformers, GEMM is critical in convolutional neural networks for image recognition (e.g., ResNet) [31], in recommendation systems for e-commerce platforms [15], and in physics simulations for climate modeling [14]. As these applications become more complex and dataintensive, optimizing GEMM performance on modern hardware architectures has become increasingly crucial for achieving efficient and scalable computations.

Conventional GEMM implementations employ data-parallel tiling of the output matrix, dividing the computation into smaller, manageable blocks that can be processed independently. These approaches often use hierarchical implementations, where the problem is broken down into progressively smaller units to match the hardware's memory hierarchy and computational capabilities. For example, a large matrix multiplication might be divided into blocks that fit into shared memory, which are further subdivided into thread-level computations that use thread-local registers [3,1].

However, as GPU cores have grown in size and complexity, a new challenge has emerged: GPU utilization [23]. Traditional tiling methods can struggle when the workload cannot be easily broken down or quantized into waves that efficiently fit onto all compute units (CUs). This issue is particularly pronounced for matrix shapes that do not align well with the GPU's architecture, leading to underutilization of computational resources and suboptimal performance. When the last round of wavefronts scheduled on the GPU only fills some of the CUs, the unused CUs must wait for the active CUs to complete millions (if not billions) of multiply-accumulate (MAC) operations before they can execute any dependent work.

Stream-K [23] addresses this challenge by introducing a novel approach to workload distribution. Instead of rigidly dividing the output matrix across the CUs, Stream-K distributes the actual work (total MAC operations), instead of the output matrix, to the CUs. This method enables finer-grained load balancing, as the streaming nature of the algorithm allows it to adapt to the available resources more dynamically. The key innovation of Stream-K lies in its ability to maintain high GPU utilization across a wide range of matrix dimensions while keeping the overhead at a constant cost. By processing the input matrices in a continuous stream, it keeps all CUs busy even when the problem size is not divisible evenly among them. This is particularly beneficial for irregular or nonsquare matrix shapes that might leave significant portions of the GPU idle under conventional tiling schemes.

Despite its promise, the optimal utilization of Stream-K presents a significant challenge. The algorithm's performance is highly sensitive to the specific Stream-K configuration used, which must be tailored to the dimensions of the matrices being multiplied (M, N, and K) and the underlying hardware characteristics. This sensitivity creates a complex optimization problem: for each GEMM operation, an appropriate Stream-K configuration must be selected to maximize performance. A configuration that performs well for one set of matrix dimensions may be suboptimal for another, necessitating a dynamic approach to configuration selection. A naive solution proposed [23] picks one Stream-K configuration for a particular GPU architecture. However, this may not be optimal in every scenario.

This paper addresses the critical need for an efficient method to select the most suitable Stream-K configuration for arbitrary GEMM problem sizes and hardware architectures. We propose Stream-K++, which significantly expands upon the original Stream-K framework by introducing four additional work schedule configurations, bringing the total to seven distinct policies. This expansion offers a more nuanced approach to workload distribution, potentially catering to a wider array of GEMM scenarios and hardware configurations. To efficiently navigate this expanded configuration space, we introduce a novel bloom-filter-based methodology for dynamic kernel selection. Implemented in Open-sieve, a new C++ header-only library, our approach utilizes a set of Bloom filters to rapidly filter out unsuitable Stream-K++ schedule configurations, narrowing down the search space for a particular GEMM problem size (M, N, K). This method successfully eliminates up to ∼95.8% of unnecessary configuration checks with a 100% true negative rate, significantly reducing selection overhead while maintaining both time and space efficiency.

Our comprehensive evaluation on AMD Instinct™ MI250X GPUs demonstrates that while Stream-K configurations may not outperform traditional approaches in most cases, they offer substantial gains of up to 43% in specific scenarios and maintain competitive performance within a small margin of 20% of the optimal configuration's performance for ∼60-97.6% of problem sizes. These findings underscore the importance of adaptive, context-aware GEMM optimization strategies. By streamlining the configuration selection process and providing a flexible framework for future optimizations, Stream-K++ not only enhances immediate GEMM performance but also paves the way for continued advancements in GPU-based matrix multiplication. In the following sections, we detail our methodology, present our experimental results, and discuss the broader implications of this work for high-performance computing on modern GPU architectures.

## 2 Related Work

Larsen et al. [20] performed matrix-matrix multiplication on GPUs by repurposing the graphics pipeline to visualize the computation as a multi-texture multiplication and blending operation. Later GPU architectures enabled highperformance GEMMs with two levels of blocking (user-programmable cache and registers) where tile sizes are chosen from extensive micro-benchmarking [9,22,27,28], and auto-tuning [11,17,21]. MAGMA GPU math library [22] is one of the early works to optimize for different GEMM problem sizes. Size thresholds based on naive handwritten rules were used for kernel selection. Subsequent GPU math libraries rely on more sophisticated code-generation and kernel-selection techniques. For example, ISAAC [28] uses machine learning to predict optimal tiling parameters. Domain Specific Languages (DSLs) like Halide [26] and TVM [10] separate the definition of the algorithm from the scheduling logic. Fireiron [16] adds constructs for scheduling data movement. Triton [29] is another popular DSL that introduces a block-based instead of thread-based programming model that automatically performs low-level optimizations like memory coalescing, shared memory management, and scheduling within CUs, and leaves the high-level logic of the parallel code to the developer.

Optimal GEMM implementations on GPUs are often found in GPU vendor libraries. The NVIDIA cuBLAS [2] (closed-source) and the AMD rocBLAS [7] (open-source) support basic linear algebra subroutines (BLAS) that include GEMMs. NVIDIA also offers CUTLASS [3], an open-source collection of CUDA C++ templates for high-performance matrix multiplication. Similarly, the AMD Composable Kernel [1,18] library provides a flexible, open-source approach to building complex GPU operations from reusable components. These libraries use hierarchical implementations of GEMMs where grids of thread blocks or wavegroups are distributed across CUs to keep the GPU occupied. Additionally, each CU has multiple resident wavegroups or blocks and each wavegroup resident on a CU may consist of one or more wavefronts or warps that are groups of 32 or 64 threads. GPUs are designed to context-switch between wavefronts with zero overhead and having multiple resident wavefronts on each CU is important to hide the memory access latency for each wavefront on the GPU. Moreover, all of these GEMM libraries rely on the custom-designed and fast Matrix Fused Multiply and Accumulate (MFMA) hardware for efficient multiply-and-accumulate (MAC) operations.

Tiling is a strategy that the GEMM libraries have adopted to reduce bandwidth usage by exploiting spatial and temporal cache locality. Input and output matrices are divided into smaller tiles or sub-matrices. Each pair of input tiles required to compute an output tile is fetched and stored in user-programmable cache (shared memory or local data share) until all the MAC operations with it are performed. In scenarios where the K dimension is very large, libraries employ a strategy known as split-K [4] where the computation is split along the K dimension and spread across multiple blocks or wavegroups to better utilize the GPU hardware.

Stream-K [23] is a recent optimization strategy that builds upon the concept of split-K for better workload balancing and hardware utilization. Stream-K does a work-centric decomposition where the total MAC operations involved in a GEMM are uniformly spread across all the CUs.

# 3 Background

## 3.1 Stream-K Algorithm

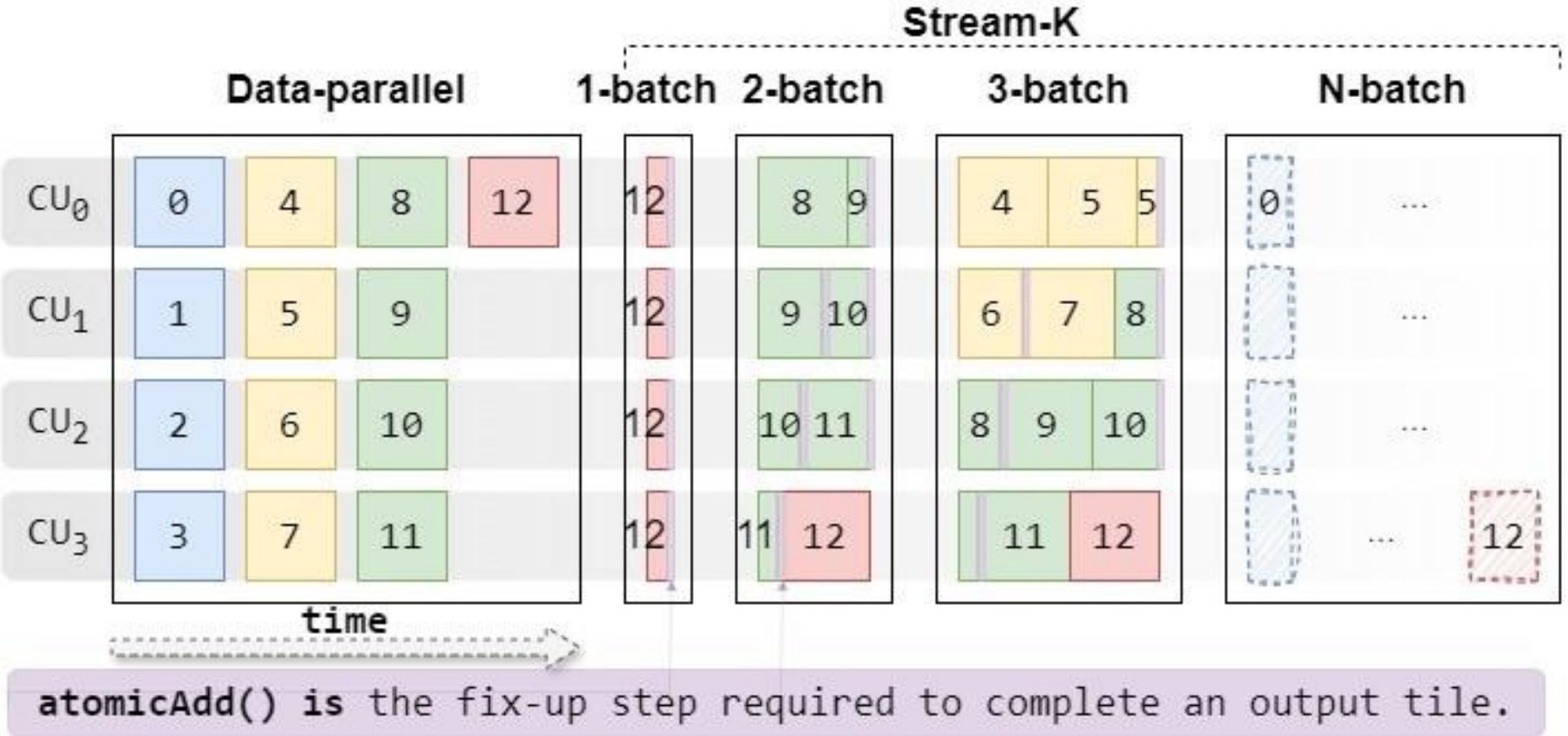

Fig.1. Stream-K distributes batches of output tile computations uniformly across CUs, allowing multiple CUs to collaboratively complete tiles using atomic operations for fixup, in contrast to traditional data-parallel GEMM approaches.

Stream-K, introduced by Osama et al. [23], represents a significant advancement in GEMM optimization for modern GPUs. This innovative strategy fundamentally restructures workload distribution by uniformly dividing the total computational work (MAC operations) across CUs, as shown in Figure 1. Stream-K surpasses the limitations of traditional approaches like split-K, which rigidly partitions output tile computations along the K dimension using a fixed factor. The core strength of Stream-K lies in its dynamic and flexible workload balancing mechanism, as illustrated in Algorithm 1. The algorithm initiates by calculating the total number of iterations required for the GEMM operation, considering the matrix dimensions and workgroup sizes (lines 2-3). It then evenly distributes these iterations among ‘g’ workgroups (line 4). Each workgroup processes its assigned range of iterations (lines 5-8), with each iteration potentially contributing to different output tiles. This approach allows for fine-grained load balancing, adapting to varying matrix sizes and hardware configurations.

The core computation occurs in the MAC_loop function (line 14), which returns a local accumulator. Each workgroup directly updates the global output matrix C using atomic add operations (line 17). This implementation of StreamK in Algorithm 1 using atomic adds for writing output to memory eliminates the need for explicit inter-workgroup synchronization and separate partial result accumulation steps. The algorithm iterates through the assigned work, computing portions of output tiles and atomically adding results to the corresponding positions in C.

Moreover, Stream-K's design inherently supports efficient deployment on modern multi-chiplet GPUs without necessitating additional tuning for parameters like grid size. This adaptability is particularly valuable in the context of evolving GPU architectures, where traditional optimization techniques may require significant adjustments for each new hardware configuration. By reimagining workload distribution and incorporating sophisticated synchronization mechanisms, Stream-K not only achieves superior load balancing but also enhances scalability and portability across diverse GPU architectures. For example, StreamK's flexible workload distribution could be particularly advantageous when running kernels on virtual cloud instances that may be allocated to specific chiplets or CUs on a multi-chiplet GPU like AMD Instinct™ MI300X. These characteristics position Stream-K as a pivotal advancement in the field of high-performance computing, particularly for large-scale matrix operations in machine learning and scientific computing applications.

**Algorithm 1** Streamk-K GEMM (using atomic adds) with grid size g

```
 1: __shared__ accum [BLK_M, BLK_N]
 2: iters_per_tile = ceil (K / BLK_K)
 3: total_iters = ceil (M / BLK_M) × ceil (N / BLK_N) × iters_per_tile
 4: iters_per_wg = ceil (total_iters/g)
        WG[x] in [g]        ⍰ Launch g workgroups 5: iter = x × iters_per_wg ⍰ Initialize
 6: iter_end = iter + iters_per_wg
7: while iter < iter_end do                         ⍰ Outer loop to process iteration
8:          tile_idx = iter / iters_per_tile
9:          tile_iter = tile_idx × iters_per_tile
10:         tile_iter_end = tile_iter + iters_per_tile
11:         local_iter = iter - tile_iter
12:         local_iter_end = min(iter_end, tile_iter_end)- tile_iter
13:             accum = MAC_loop (tile_idx, local_iter, local_iter_end) ⍰ MAC iterations
    for this tile
14:         for i in 0 to BLK_M do
15:             for j in 0 to BLK_N do
16:                     atomic_add(C [tile_idx.m + i, tile_idx.n + j], accum [i, j])
17:             end for
18:         end for
19:         iter = tile_iter_end
20: end while
21: join
```

### 3.2 Challenges in Stream-K Configuration Selection

Osama et al. [23] introduce three different Stream-K schedules in their work: basic Stream-K, data-parallel followed by one-batch Stream-K, and two-batch Stream-K followed by data-parallel. Each "batch" or "round" in this context represents a single iteration (as in line 8 on Algorithm 1 and in Figure 1) for all GPU wavefronts, processing their allocated workload. In the basic or allStream-K configuration, as shown in Algorithm 1, the total workload is equally distributed among all the workgroups. Here, the workgroups do not perform the conventional output tile-based (data-parallel) work.

In the second configuration of data-parallel followed by one-batch Streamk, all the workgroups perform multiple batches or rounds of the conventional data-parallel GEMM computation, followed by one batch of Stream-K computation on whatever work that remains. However, this fails to hide the latencies of atomic adds to the output. For instance, the atomic adds into the shared output tile can take thousands of clock cycles, potentially creating a performance bottleneck at the end of the computation. To address these limitations, the authors propose a third policy as an improvement over the second one. In this approach, two batches of Stream-K work precede the conventional data-parallel computation. Initiating with Stream-K batches facilitates the overlap of the atomic write latencies with the data-parallel computation.

However, the authors leave two critical areas unexplored: the potential benefits of extending the Stream-K methodology beyond two batches, and a systematic methodology for selecting optimal Stream-K configurations across various GEMM sizes and hardware architectures.

Stream-K++ significantly enhances and expands upon the original StreamK framework, introducing a comprehensive suite of seven distinct scheduling policies. These policies range from the all-Stream-K approach to a series of hybrid configurations, systematically increasing the number of Stream-K batches (from zero to six) along with the configuration for the conventional data-parallel computation. This expanded set of policies offers a more nuanced and flexible approach to workload distribution, potentially catering to a wider array of GEMM scenarios and hardware configurations. Moreover, we introduce a novel bloom-filter-based methodology for dynamically selecting the optimal scheduling policy. This innovative selection mechanism efficiently chooses between the seven Stream-K++ policies, adapting to the specific characteristics of each GEMM operation and hardware environment. By combining an extended range of scheduling options with an intelligent selection mechanism, Stream-K++ represents a significant advancement in GPU-based matrix multiplication optimization, promising improved performance across diverse computational.

## 4 Our Approach

### 4.1 Stream-K++: Scheduling Policy

Our implementation of the Stream-K++'s scheduling policy begins with the launch of a persistent kernel. The grid size for this kernel is carefully optimized to maximize GPU utilization, taking into account both the compilerdetermined occupancy and the number of CUs available on the target GPU.
This initial step ensures efficient resource allocation from the outset. Following the kernel launch, we employ a dynamic workload distribution strategy based on user-defined Stream-K++ configurations. This approach allows for flexible partitioning of work across CUs, adapting to various problem sizes and GPU architectures.

The workload is then assigned to wavefronts in two distinct batch types: Stream-K batches (ranging from zero to six based on input from the user or ckProfiler) and data-parallel batches. This hybrid assignment leverages the loadbalancing strengths of Stream-K while maintaining the high throughput capabilities of traditional data-parallel execution. A key feature of our implementation is the strategic overlap of execution, where the latencies inherent in the atomic adds of Stream-K batches are effectively masked by the concurrent processing of data-parallel batches. This technique minimizes idle time and maximizes computational efficiency, potentially leading to significant performance gains across a wide spectrum of matrix dimensions and shapes.

4.2 Bloom Filter Design for Efficient Stream-K++ Policy Selection The integration of additional Stream-K++ policies into ckProfiler significantly expands the search space for optimal GEMM configurations, potentially leading to prohibitive increases in tuning time for our FP16 GEMM benchmark. To address this challenge, we have developed a novel approach utilizing space and time-efficient Bloom filters for each Stream-K++ policy. This innovative design ensures a 100% true negative rate when determining whether a specific GEMM size is associated with any of the seven filters corresponding to the Stream-K++ policies.

Bloom filters, renowned for their efficiency in set membership queries, form the cornerstone of our solution. We employ the popular Murmur hash implementation, mmh3 [6], to generate unique keys from the problem size parameters (M, N, K), which are then used to query the Bloom filters. Our implementation, Open-sieve, is a C++ header-only library that maps winning Stream-K++ configurations for GEMM problem sizes to their corresponding Bloom filters. We utilize 7 distinct hash functions, one for each filter, and configure the Bloom filters to accommodate 10,000 problem sizes each, minimizing false positives for our 923 GEMM sizes while maintaining scalability for future expansions.

As a one-time preprocessing step, Open-sieve analyzes all 923 problem sizes in our benchmark suite, encoding the winning configurations into a compact C++ header file. This file, containing approximately 1 byte of information per problem size, serves as a highly efficient lookup table. When queried with a new GEMM problem size, the system rapidly checks the Bloom filters to determine if (M, N, K) corresponds to any Stream-K++ policies. Our optimized implementation achieves a remarkably low query time of 0.4 microseconds per lookup on a single CPU thread.

Our experimental results demonstrate the exceptional efficiency of this approach. The Bloom filter-based checks eliminate up to ~95.8% of the additional policy evaluations that would otherwise be required when using ckProfiler. This significant reduction in computational overhead not only accelerates the tuning process but also enhances scalability. Furthermore, our method readily accommodates additional GEMM problem sizes and Stream-K++ policies, and even extends to other tuning parameters beyond Stream-K++, all while maintaining exceptional space and time efficiency.

This solution represents a significant advancement in GEMM optimization techniques, offering a scalable and efficient approach to navigating the complex landscape of Stream-K++ policy selection.

## 5 Evaluation

### 5.1 Experiment Setup

Our experiments were conducted on an AMD Instinct™ MI250X GPU operating in its default Single Process Execution (SPX) mode, utilizing all 104 CUs on a single chiplet for computation. We designed a comprehensive FP16 benchmark suite comprising 923 unique GEMM problem sizes, with dimensions varied in powers of 2 within the following ranges: M from 1 to 8192, N from 64 to 8192, and K from 16 to 65536. We chose FP16 because it has emerged as a de facto standard for many popular inference engines [5] [24] [8]. While the GEMM problem ranges are informed by industry trends, they were generalized to maintain confidentiality and ensure broad applicability.

To evaluate GEMM performance, we employed the AMD Composable Kernel library [1,18], specifically utilizing the ckProfiler tool. ckProfiler is a tool that systematically evaluates various GEMM instances, aiming to determine the GEMM instance with the optimal wavegroup configurations and Stream-K++ policy for maximizing performance. For benchmarking each configuration, we perform fifty GPU kernel launches to warm the GPU up to a steady state and then measure performance as an average of another fifty kernel launches.

For in-depth performance analysis of the optimal kernels identified by ckProfiler, we used rocProf v6.1, which is an AMD GPU profiler included in the AMD ROCm™ 6.1 software stack. This profiling tool provided detailed insights into kernel execution characteristics and resource utilization. We used a 64-core AMD EPYC™ 7713 processor for CPU-related tasks.

### 5.2 GEMM Performance Analysis

Our comprehensive analysis reveals nuanced insights into the performance dynamics of Stream-K++ policies compared to conventional data-parallel approaches. As illustrated in Figure 2, the data-parallel policy emerges as the optimal configuration for ∼87% of the GEMM problem sizes. However, this initial observation belies the more complex performance landscape of Stream-K-based schedules.

A key finding emerges when we examine performance within varying tolerance thresholds. As we expand the acceptable performance slow-down margin from 5% to 20% relative to the data-parallel or no-stream-K baseline, we observe a dramatic increase in the prevalence of Stream-K-based winners—from approximately ∼60% to ∼97.6%. This significant shift underscores that while Stream-K configurations may not win most of the time, they consistently deliver competitive performance, often within a narrow margin of the data-parallel approach.

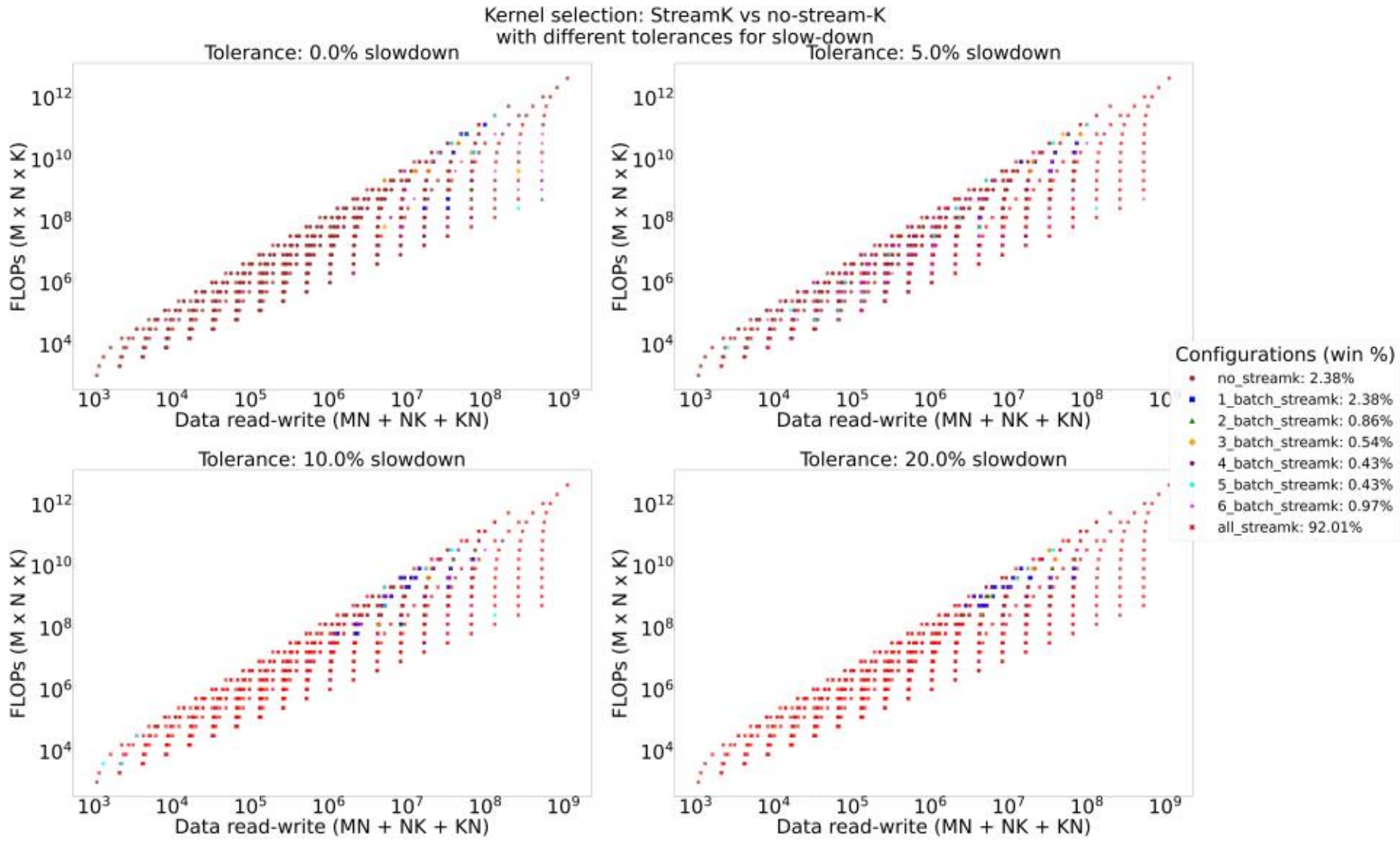

Fig.2. A significant fraction of the winning kernel configurations in Stream-K++ are from Stream-K-based work schedules when we allow for some tolerance in slowdown with respect to the data-parallel schedule.

The apparent discrepancy between the ∼13% optimal occurrence rate of Stream-K configurations and their broader competitiveness warrants further investigation. To this end, Figure 3 provides crucial insights through a detailed violin plot analysis. This visualization contrasts the performance gains of winning configurations for both Stream-K and data-parallel approaches relative to their runners-up.

Stream-K schedules exhibit a distinctly asymmetric performance distribution. The mean gain (denoted by a green dot) substantially exceeds the median (indicated by a dashed line), revealing a right-skewed distribution characterized by several high-impact cases. This stands in stark contrast to the more symmetrical distribution observed for data-parallel configurations. Notably, the extended upper tail of the Stream-K violin plot highlights instances of exceptional performance enhancement, with some cases showing improvements of over ∼40% compared to the runner-up.

These findings underscore the strategic importance of incorporating StreamK schedules in GEMM optimization frameworks. Despite their less frequent occurrence as globally optimal solutions, Stream-K configurations demonstrate the potential for substantial performance improvements in specific scenarios. This potential for high-impact optimization, coupled with their broad competitiveness across problem sizes, reinforces the value of the expanded policy set introduced in Stream-K++.

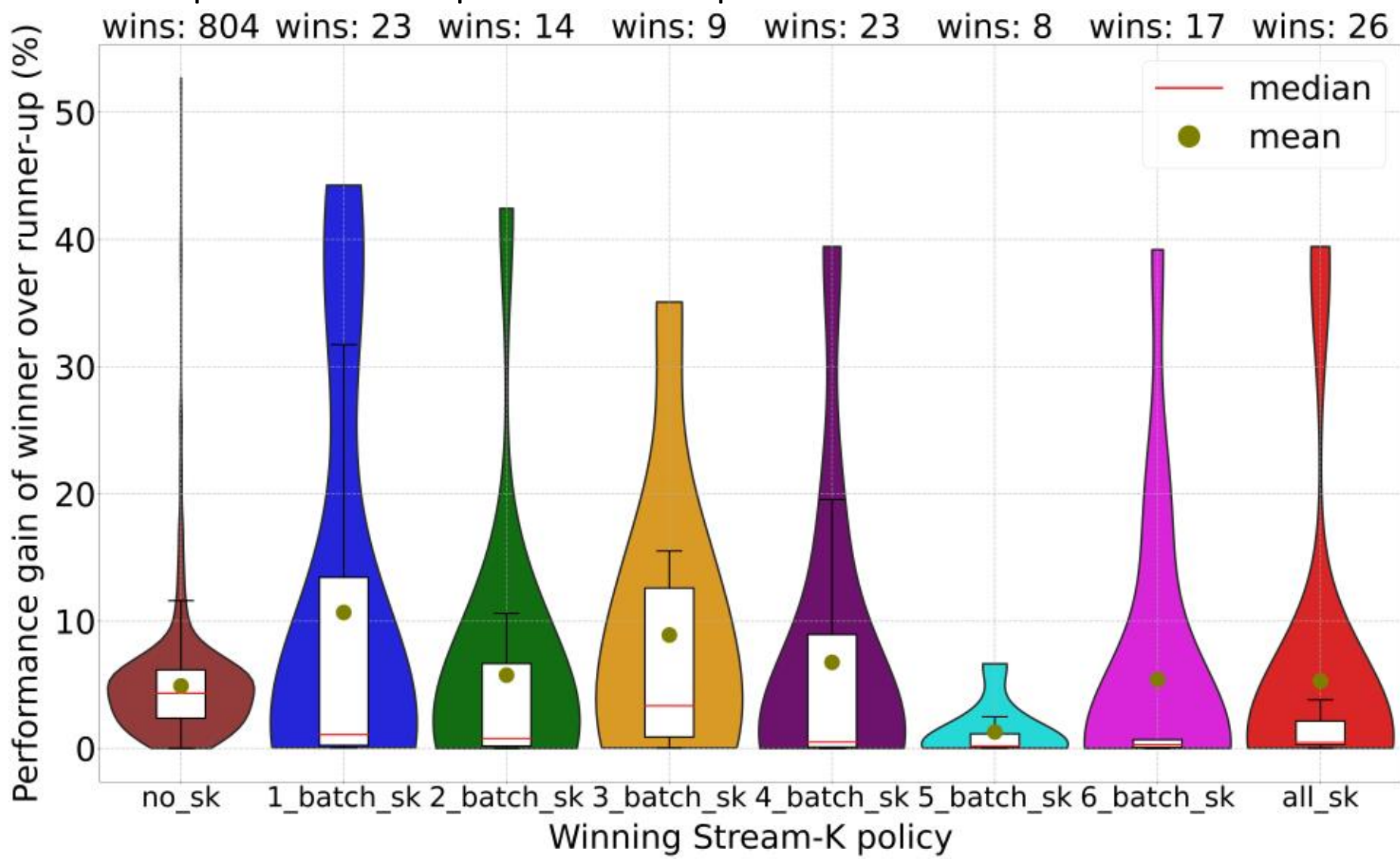

Fig.3. Although just 13% of the winning kernel configurations are from Stream-Kbased schedules, the gains are significant. The median performance gain for the Streamk schedules over the runner-up configuration is significantly higher than the mean.

Our results not only justify the inclusion of all Stream-K configurations but also emphasize the necessity of sophisticated selection mechanisms to leverage these policies effectively. The significant performance gains observed in select cases, combined with the overall competitiveness of Stream-K schedules, highlight the importance of a nuanced, adaptive approach to GEMM optimization that can identify and exploit these high-impact scenarios.

## 5.3 Discussion

Our analysis reveals important insights about Stream-K performance. While data-parallel policies are optimal for the majority of problem sizes, StreamK configurations demonstrate competitive performance within narrow margins and exceptional gains in select scenarios. The asymmetric performance distribution of Stream-K schedules, with improvements exceeding 40% in some cases, underscores their strategic importance in GEMM optimization frameworks.

We identify two primary challenges in Stream-K implementation. Firstly, despite the multi-batch schedules in Stream-K++ helping to mitigate some latencies associated with atomic adds, this remains problematic for large K dimensions. While atomic adds simplify the algorithm by eliminating explicit inter-workgroup synchronization and separate partial result accumulation, an alternative approach using parallel reduction to combine partial results from wavegroups could potentially yield superior results. However, the effect of interworkgroup synchronization for parallel reduction on performance would need to be investigated.

Secondly, we observed increased L1 cache misses as a limitation in certain cases, with Stream-K-based schedules suffering up to ∼30% lower L1 cache hits compared to no-Stream-K in scenarios where the latter prevails. This finding emphasizes the need to incorporate cache locality considerations when allocating work to CUs, a factor that will become increasingly critical for multi-chiplet GPUs with disaggregated caches.

Our current study focuses on FP16 GEMMs on the MI250X GPU. Future work will expand to other precisions and mixed precision (FP32, FP16, FP6, FP4), profiling energy efficiency of Stream-K++, exploring additional tuning parameters beyond Stream-K policies, and diverse GPU architectures and form factors. This expansion will provide a more comprehensive understanding of Stream-K's applicability and performance across various computational contexts.

The flexibility of our Bloom filter approach in Open-sieve is a key strength, allowing for easy modification to accommodate new problem sizes, policies, or tuning parameters. New key-value entries corresponding to additional GEMM problem sizes can be swiftly integrated into the Bloom filters. Moreover, new policies or tuning parameters can be incorporated through the creation of new Bloom filters. This extensibility positions Stream-K++ as a versatile framework for ongoing GEMM optimization research, adaptable to evolving hardware landscapes and computational demands.

## 6 Conclusion

Stream-K++ represents a significant advancement in GPU-based matrix multiplication optimizations for domains including AI. By expanding Stream-K policies and introducing an efficient Bloom filter-based selection mechanism, we have enhanced the practicality and performance of Stream-K across diverse GEMM operations. Our results demonstrate that while Stream-K configurations may not outperform traditional approaches in most cases, they offer substantial gains in specific scenarios and maintain competitive performance overall.

The asymmetric performance distribution of Stream-K schedules reveals their potential for exceptional optimization in certain cases, justifying their inclusion in GEMM frameworks despite less frequent occurrences as globally optimal solutions. This research not only improves immediate GEMM performance but also establishes a flexible foundation for future optimizations, particularly as GPU architectures continue to evolve.

The insights gained from this study highlight the importance of nuanced, adaptive approaches to GEMM optimization that can identify and exploit highimpact scenarios. As we continue to refine these techniques, the potential for significant performance enhancements in critical computational tasks across AI, scientific computing, and beyond becomes increasingly apparent.